\documentclass{svjour3}                     
\smartqed  

\usepackage[authoryear]{natbib}
\usepackage[centertags]{amsmath}
\usepackage{bm}

\newcommand{\vvec}{\textrm{vec}}
\newcommand{\tr}{\textrm{tr}}
\newcommand{\Es}{\textrm{E}}

\begin{document}

\title{Improved maximum likelihood estimators in a heteroskedastic errors-in-variables model}

\author{Alexandre G.~Patriota\and Artur J.~Lemonte \and  Heleno Bolfarine}

\institute{A.~G.~Patriota\and A.~J.~Lemonte \and  H.~Bolfarine\at
Departamento de Estat\'istica, Universidade de S\~ao Paulo, Rua do Mat\~ao, 1010,
S\~ao Paulo/SP, 05508-090, Brazil\\
                Fax: +55\ 11\ 38144135\\
\email{patriota.alexandre@gmail.com}
}
\date{Received: date / Accepted: date}
\journalname{Statistical Papers}

\maketitle

\begin{abstract}
This paper develops a bias correction scheme for a
multivariate heteroskedastic errors-in-variables model. The applicability
of this model is justified in areas such as astrophysics, epidemiology and analytical
chemistry, where the variables are subject to measurement errors and the variances vary
with the observations. We conduct Monte Carlo simulations to investigate the performance of the
corrected estimators. The numerical results show that the bias correction scheme
yields nearly unbiased estimates.
We also give an application to a real data set.
\end{abstract}

\section{Introduction}

Heteroskedastic errors-in-variables (or measurement error) models have been extensively
studied in the statistical literature and widely applied in astrophysics (to explain
relationships between black hole masses and some variates of luminosities), epidemiology
(to model the cardiovascular event with its risk factors), analytical chemistry (to compare
different types of measurement instruments).  The applicability of this model abound mainly
in the astronomy literature where all quantities are subject to measurement errors \citep{Akritas1996}.

It is well-known that, when the measurement errors are ignored in the estimation process,
the maximum-likelihood estimators (MLEs) become inconsistent. More specifically, the
estimation of the slope parameter of a simple linear model is attenuated \citep{Fuller}. 
When variables are subject to
measurement errors, a special inference treatment must be carried out for the model
parameters in order to avoid inconsistent estimators. Usually, a measurement equation
is added to the model to capture the measurement error effect and then the MLEs from
this approach are consistent, efficient and asymptotically normally distributed.
A careful and deep exposition on the inferential process in errors-in-variables models can be
seen in \cite{Fuller} and  the references therein.

Although consistent, asymptotically efficient and asymptotically normally dis\-tri\-bu\-ted,
the MLEs are oftentimes biased and point inference can be misleading. This is not a serious
problem for relatively large sample sizes, since bias is typically
of order $\mathcal{O}(n^{-1})$, while the asymptotic standard errors
are of order $\mathcal{O}(n^{-1/2})$. However, for small or even moderate
values of the sample size $n$, bias can constitute a problem. Bias adjustment
has been extensively studied in the statistical literature. For example,
\cite{CTWei(1986)}, \cite{Cord93}, \cite{CorVasc1997}, \cite{Vasccor1997} and,
more recently, \cite{Cordeiro2008}.
Additionally, \cite{PatriotaLemonte2009} obtained general matrix formulae
for the second-order biases of the maximum-likelihood estimators in a very general
model which includes all previous works aforementioned. The model presented
by the authors considers that the mean vector and the variance-covariance
matrix of the observed variable have parameters in common.
This approach includes the heteroskedastic measurement error model that we are
going to study in this paper.

The main goal of this article is to define bias-corrected estimators
using the general second-order bias expression derived in \cite{PatriotaLemonte2009} assuming
that the model defined by (\ref{Mainmodel}) and (\ref{model}) holds.
Additionally, we compare the performance of bias-corrected
estimators with the MLEs in small samples via Monte Carlo simulations.
The numerical results show that the bias correction is effective in small samples and leads to
estimates that are nearly unbiased and display superior finite-sample behavior.

The rest of the paper is as follows. Section~\ref{modelo-heteros} presents the
multivariate heteroskedastic errors-in-variables model. Using general results from
\cite{PatriotaLemonte2009}, we derive in Section~\ref{bias-theta}
the second-order biases of the MLEs of the parameters.
The result is used to define bias-corrected estimates. In Section~\ref{mu-sigma-vies}
the $O(n^{-1})$ biases of the estimates $\widehat{\bm{\mu}}_{i}$ and $\widehat{\bm{\Sigma}}_{i}$
are given. Monte Carlo simulation results are presented and discussed in Section~\ref{simulations}.
Section~\ref{application} gives an application. Finally, concluding remarks are offered
in Section~\ref{conclusion}.

\section{The model}\label{modelo-heteros}

The multivariate model assumed throughout this paper is 
\begin{equation}\label{Mainmodel}
\bm{y}_i = \bm{\beta}_0 + \bm{\beta}_1 \bm{x}_i +\bm{q}_i,\quad i=1,\ldots, n,
\end{equation} 
where $\bm{y}_i$ is a ($v\times 1$)
latent response vector, $\bm{x}_i$ is a ($m\times 1$) latent vector
of covariates, $\bm{\beta}_0$ is a ($v\times 1$) vector of intercepts,
$\bm{\beta}_1$ is a ($v \times m$) matrix, the elements of which are inclinations
and $\bm{q}_i$ is the equation error having a multivariate normal distribution
with mean zero and covariance-variance matrix $\bm{\Sigma}_{\bm{q}}$. The variables
$\bm{y}_i$ and $\bm{x}_i$ are not directly observed, instead surrogate variables
$\bm{Y}_i$ and $\bm{X}_i$ are measured with the following additive structure:
\begin{equation}\label{model}
\bm{Y}_i = \bm{y}_i + \bm{\eta}_{\bm{y}_{i}} \quad \mbox{and} \quad \bm{X}_i =
\bm{x}_i + \bm{\eta}_{\bm{x}_{i}}.
\end{equation}
The errors $\bm{\eta}_{\bm{y}_{i}}$ and $\bm{\eta}_{\bm{x}_{i}}$ are assumed
to follow a normal distribution given by
\[
\begin{pmatrix}
 \bm{\eta}_{\bm{y}_{i}}\\
 \bm{\eta}_{\bm{x}_{i}}\\
\end{pmatrix}
\stackrel{ind}{\sim}\mathcal{N}_{v+m}
\begin{bmatrix}
\begin{pmatrix}
 \bm{0}\\
 \bm{0}\\
\end{pmatrix},
\begin{pmatrix}
 \bm{\tau}_{\bm{y}_{i}} & \bm{0}        \\
  \bm{0}              & \bm{\tau}_{\bm{x}_{i}}\\
\end{pmatrix}\end{bmatrix},
\]
where ``$\stackrel{ind}{\sim}$'' means ``independently distributed as'' and the 
covariance-variance matrices $\bm{\tau}_{\bm{y}_{i}}$ and $\bm{\tau}_{\bm{x}_{i}}$ are
assumed to be known for all $i=1,\ldots,n$. These matrices may be attained, for example, through
an analytical treatment of the data collection mechanism, 
replications, machine precision, etc.

Model (\ref{model}) has equation errors for all lines, i.e.,
$\bm{y}_{i}$ and $\bm{x}_i$ are not perfectly related. These
equation errors are justified by the influence of other factors than
$\bm{x}_i$ in the variation of $\bm{y}_i$. It is very reasonable to
consider equation errors in (\ref{Mainmodel}) to capture extra variability,
since the variances $\bm{\tau}_{\bm{y}_{i}}$ are fixed and whether some other factor
affects the variation of $\bm{y}_i$, the estimation of the line parameters
will be clearly affected. Supposing that
$\bm{x}_i \stackrel{iid}{\sim} \mathcal{N}_m(\bm{\mu}_{\bm{x}},\bm{\Sigma}_{\bm{x}})$,
where ``$\stackrel{iid}{\sim}$'' means ``independent and identically distributed as'',
and considering that the model errors ($\bm{q}_i$, $\bm{\eta}_{\bm{y}_{i}}$ and $\bm{\eta}_{\bm{x}_{i}}$)
and $\bm{x}_i$ are independent, we have that the joint distribution of the
observed variables can be expressed as
\begin{equation}\label{model-Dist}
\begin{pmatrix}
\bm{Y}_i\\
\bm{X}_i\end{pmatrix}
\stackrel{ind}{\sim}\mathcal{N}_{v+m}
\begin{bmatrix}
\begin{pmatrix}
\bm{\beta}_0 + \bm{\beta}_1 \bm{\mu}_{\bm{x}}\\
\bm{\mu_x}
\end{pmatrix},
\begin{pmatrix}
\bm{\beta}_1\bm{\Sigma}_{\bm{x}}\bm{\beta}_1^{\top}+ \bm{\Sigma}_{\bm{q}} +
\bm{\tau}_{\bm{y}_{i}} & \bm{\beta}_1 \bm{\Sigma}_{\bm{x}}\\
\bm{\Sigma}_{\bm{x}}\bm{\beta}_1^{\top} &\bm{\Sigma}_{\bm{x}} + \bm{\tau}_{\bm{x}_{i}}
\end{pmatrix}
\end{bmatrix}.
\end{equation}

Note that in~(\ref{model-Dist}), the mean vector and the covariance-variance
matrix of observed variables have the matrix $\bm{\beta}_1$ in common, i.e., they
share $mv$ parameters. \cite{Kulathinal2002} study the univariate case (when $v=1$ and $m=1$)
and propose an EM (Expectation and Maximization) algorithm to obtain MLEs for
model parameters. In addition, they derived the asymptotic variance of the
MLE of the inclination parameter making it possible to build hypotheses testing
of it. Also, \cite{deCastro2008} derive the observed and expected Fisher information
and conduct some simulation studies to investigate the behavior of the likelihood ratio,
score, Wald and $C(\alpha)$ statistics for testing hypothesis of the parameters
 and \cite{Patriota} study the asymptotic properties of method-of-moments estimators 
in the univariate model proposed by \cite{Kulathinal2002}.
 Model~(\ref{model}) is a multivariate version of the model proposed by
\cite{Kulathinal2002}.

\section{Second-order bias of $\widehat{\bm{\theta}}$}\label{bias-theta}

In order to follow the same scheme adopted by \cite{PatriotaLemonte2009},
define the vector of parameters $\bm{\theta} = (\bm{\beta}_0^{\top}, \vvec(\bm{\beta}_1)^{\top},
\bm{\mu}_{\bm{x}}^{\top}, \mbox{vech}(\bm{\Sigma}_{\bm{x}})^{\top},
\mbox{vech}(\bm{\Sigma}_{\bm{q}})^{\top})^{\top}$,
where vec$(\cdot)$ is the vec operator,
which transforms a matrix into a vector by stacking the columns of the matrix and
vech$(\cdot)$ is the vech operator, which transforms a symmetric matrix
into a vector by stacking the on or above diagonal elements.
Also, consider $\bm{Z}_{i} = (\bm{Y}_{i}^{\top}, \bm{X}_{i}^{\top})^{\top}$
and the mean and covariance-variance function as
\[
\bm{\mu}_i (\bm{\theta})= {\bm{\beta}_0 + \bm{\beta}_1\bm{\mu}_{\bm{x}} \choose \bm{\mu}_{\bm{x}}}
\quad \mbox{and} \quad\bm{\Sigma}_i(\bm{\theta}) =
\begin{pmatrix}
\bm{\beta}_1\bm{\Sigma}_{\bm{x}}\bm{\beta}_1^{\top} + \bm{\Sigma}_{\bm{q}} + \bm{\tau}_{\bm{y}_{i}}
& \bm{\beta}_1\bm{\Sigma}_{\bm{x}} \\
\bm{\Sigma}_{\bm{x}}\bm{\beta}_1^{\top} & \bm{\Sigma}_{\bm{x}}+ \bm{\tau}_{\bm{x}_{i}}
\end{pmatrix},
\]
respectively.

Moreover, to simplify notation, define the quantities $\bm{Z}= \vvec(\bm{Z}_1,\ldots,\bm{Z}_n)$,
$\bm{\mu} = \vvec(\bm{\mu}_1(\bm{\theta}),\ldots,\bm{\mu}_n(\bm{\theta}))$,
$\bm{\Sigma} = \mbox{block--diag}\{\bm{\Sigma}_1(\bm{\theta}), \ldots,\bm{\Sigma}_n(\bm{\theta})\}$
and $\bm{u} = \bm{Z} - \bm{\mu}$.
The log-likelihood function for the vector parameter $\bm{\theta}$ from a
random sample, except for constants, can be expressed as
\begin{equation}\label{Ver}
\ell(\bm{\theta}) = -\dfrac{1}{2} \log{|\bm{\Sigma}|}
-\dfrac{1}{2}\tr\{\bm{\Sigma}^{-1}\bm{u}\bm{u}^{\top}\}.
\end{equation}
Additionally, for the purpose of computing the score function,
the Fisher information and the second-order biases, 
also define
\[
\bm{a}_r = \frac{\partial \bm{\mu}}{\partial \theta_{r}}, \quad
\bm{a}_{sr} = \frac{\partial^2 \bm{\mu}}{\partial \theta_{s} \partial\theta_{r}}, \quad
\bm{C}_{r} = \frac{\partial \bm{\Sigma}}{\partial \theta_{r}},\quad
\bm{C}_{sr} = \frac{\partial \bm{C}_{r}}{\partial \theta_{s}},\quad
\bm{A}_{r} =  -\bm{\Sigma}^{-1} \bm{C}_r\bm{\Sigma}^{-1}
\]
and
\begin{align}\label{expres}
\begin{split}
\bm{F}_{\bm{\beta}_0}^{(r)}&=\dfrac{\partial \bm{\beta}_0}{\partial \theta_r}, \quad
\bm{F}_{\bm{\beta}_1}^{(s)} = \dfrac{\partial \bm{\beta}_1}{\partial \theta_s}, \quad 
\bm{F}_{\bm{\mu}_{\bm{x}}}^{(s)} =\dfrac{\partial \bm{\mu}_{\bm{x}}}{\partial \theta_s},\\
& \bm{F}_{\bm{\Sigma}_{\bm{x}}}^{(s)} = \dfrac{\partial \bm{\Sigma}_{\bm{x}}}{\partial \theta_s}
\quad \mbox{and}\quad
\bm{F}_{\bm{\Sigma}_{\bm{q}}}^{(s)} = \dfrac{\partial\bm{\Sigma}_{\bm{q}}}{\partial \theta_s},
\end{split}
\end{align}
with $r,s=1,2,\ldots,p$, where $p$ is the dimension of $\bm{\theta}$. The quantities (\ref{expres}) are
vectors or matrices of zeros with a unit in the position referring to the $s^{{\rm th}}$ element of $\bm{\theta}$.
Let $\widetilde{\bm{D}} = (\bm{a}_{\bm{\beta}_0}, \bm{a}_{\bm{\beta}_1},\bm{a}_{\bm{\mu}_{\bm{x}}},\bm{0}, \bm{0})$
and $\widetilde{\bm{V}} = (\bm{0},\bm{C}_{\bm{\beta}_1},\bm{0},\bm{C}_{\bm{\Sigma}_{\bm{x}}},
\bm{C}_{\bm{\Sigma}_{\bm{q}}})$, with $\bm{a}_{\bm{\beta}_0} = (\bm{a}_{1}, \bm{a}_{2},\ldots,\bm{a}_{v})$,
$\bm{a}_{\bm{\beta}_1} = (\bm{a}_{v+1}, \ldots,\bm{a}_{v(m+1)})$,
$\bm{a}_{\bm{\mu}_{\bm{x}}} = (\bm{a}_{v(m+1) + 1}, \ldots,\bm{a}_{v(m+1) + m})$,
$\bm{C}_{\bm{\beta}_1} = \bigl(\vvec(\bm{C}_{v+1}), \ldots,\vvec(\bm{C}_{v(m+1)})\bigr)$,
$\bm{C}_{\bm{\Sigma}_{\bm{x}}} = \bigl(\vvec(\bm{C}_{(v+1)(m+1)}), \ldots,\vvec(\bm{C}_{p'})\bigr)$ and 
$\bm{C}_{\bm{\Sigma}_{\bm{q}}} = \bigl(\vvec(\bm{C}_{p'+1}),
\ldots,\vvec(\bm{C}_{p})\bigr)$, where $p'= v(m+1)+m+m(m+1)/2$. 

The first derivative of (\ref{Ver}) with respect to the $r^{{\rm th}}$
element of $\bm{\theta}$ is 
\begin{equation}\label{score}
U_{r} = \dfrac{1}{2} \tr\{\bm{A}_{r}(\bm{\Sigma} - \bm{u}\bm{u}^{\top})\} +
\tr\{\bm{\Sigma}^{-1}\bm{a}_{r} \bm{u}^{\top}\};
\end{equation}
the expectation of the derivative of (\ref{score}) with respect to the
$s^{{\rm th}}$ element of $\bm{\theta}$ is given by
\begin{equation*}\label{krs}
\kappa_{sr} = \frac{1}{2}\tr\{\bm{A}_{r}\bm{C}_{s}\} -  \bm{a}_{s}^{\top}\bm{\Sigma}^{-1}\bm{a}_{r}.
\end{equation*}
Under general regularity conditions (Cox and Hinkley, 1974, Ch.~9),
$-\kappa_{sr}$ is the $(s,r)^{{\rm th}}$ element of the expected Fisher information.
The score function and the expected Fisher information are given, respectively, by 
$\bm{U}_{\bm{\theta}} = \widetilde{\bm{D}}^{\top}\bm{\Sigma}^{-1}\bm{u}
-\frac{1}{2}\widetilde{\bm{V}}^{\top}\widetilde{\bm{\Sigma}}^{-1}\vvec(\bm{\Sigma} - \bm{u}\bm{u}^{\top})$
and $\bm{K}_{\bm{\theta}} = \widetilde{\bm{D}}^{\top}\bm{\Sigma}^{-1} \widetilde{\bm{D}}
+\frac{1}{2}\widetilde{\bm{V}}^{\top}\widetilde{\bm{\Sigma}}^{-1}\widetilde{\bm{V}}$,
with $\widetilde{\bm{\Sigma}} = \bm{\Sigma} \otimes \bm{\Sigma}$ and $\otimes$ is the Kronecker product. 
Defining
\[
\widetilde{\bm{u}} =
\begin{pmatrix}
\bm{u}\\
-\vvec(\bm{\Sigma} - \bm{u}\bm{u}^{\top})
\end{pmatrix}, \quad
\widetilde{\bm{F}} = 
\begin{pmatrix}
\widetilde{\bm{D}}\\
\widetilde{\bm{V}}
\end{pmatrix}\quad \mbox{and}\quad
\widetilde{\bm{H}} =
\begin{pmatrix}
\bm{\Sigma} & \bm{0}\\
\bm{0} & 2\widetilde{\bm{\Sigma}}
\end{pmatrix}^{-1},
\]
we can write the score function and the Fisher information in a short form as
\[
\bm{U}_{\bm{\theta}} = \widetilde{\bm{F}}^{\top}\widetilde{\bm{H}}\widetilde{\bm{u}}
\quad \mbox{and} \quad 
\bm{K}_{\bm{\theta}} = \widetilde{\bm{F}}^{\top}\widetilde{\bm{H}}\widetilde{\bm{F}}.
\]

The Fisher scoring method can be used to
estimate $\bm{\theta}$ iteratively solving the equation
\begin{equation}\label{Fisher-Scoring}      
\bm{\theta}^{(m+1)} = (\widetilde{\bm{F}}^{(m)\top}\widetilde{\bm{H}}^{(m)}\widetilde{\bm{F}}^{(m)})^{-1}
\widetilde{\bm{F}}^{(m)\top}\widetilde{\bm{H}}^{(m)}\widetilde{\bm{u}}^{*(m)},\quad
m = 0, 1, 2,\ldots,
\end{equation}
where $\widetilde{\bm{u}}^{*(m)} = \widetilde{\bm{F}}^{(m)}\bm{\theta}^{(m)}
+ \widetilde{\bm{u}}^{(m)}$. Each loop, through the iterative scheme~(\ref{Fisher-Scoring}),
consists of an iterative re-weighted least squares algorithm to optimize the log-likelihood~(\ref{Ver}).
Using equation~(\ref{Fisher-Scoring}) and any software
({\tt MAPLE}, {\tt MATLAB}, {\tt Ox}, {\tt R}, {\tt SAS}) with a
weighted linear regression routine one can compute the MLE, $\widehat{\bm{\theta}}$, iteratively.
Initial approximation $\bm{\theta}^{(0)}$
for the iterative algorithm is used to evaluate
$\widetilde{\bm{F}}^{(0)}$, $\widetilde{\bm{H}}^{(0)}$ and $\widetilde{\bm{u}}^{*(0)}$ from which these
equations can be used to obtain the next estimate $\bm{\theta}^{(1)}$.
This new value can update $\widetilde{\bm{F}}$, $\widetilde{\bm{H}}$ and $\widetilde{\bm{u}}^{*}$
and so the iterations continue until convergence is achieved.

The general matrix formulae derived by \cite{PatriotaLemonte2009} for $n^{-1}$ bias vector
$\bm{B}(\widehat{\bm{\theta}})$ of $\widehat{\bm{\theta}}$ is given by
\begin{equation}\label{BIAS}
\bm{B}(\widehat{\bm{\theta}}) = (\widetilde{\bm{F}}^{\top}\widetilde{\bm{H}}\widetilde{\bm{F}})^{-1}
\widetilde{\bm{F}}^{\top}\widetilde{\bm{H}}\widetilde{\bm{\xi}},
\end{equation}
where $\widetilde{\bm{\xi}} = (\bm{\Phi}_1,\ldots,\bm{\Phi}_p)\vvec\{(\widetilde{\bm{F}}^{\top}
\widetilde{\bm{H}}\widetilde{\bm{F}})^{-1}\}$ and
$\bm{\Phi}_{r} = -\frac{1}{2}(\bm{G}_{r} + \bm{J}_r)$, $r = 1, 2, \ldots, p$, with 
\[
\bm{G}_{r}=
\begin{bmatrix}
\bm{a}_{1r} & \cdots & \bm{a}_{pr}\\
\vvec(\bm{C}_{1r})& \cdots &    \vvec(\bm{C}_{pr})
\end{bmatrix}\quad{\rm and}\quad
\bm{J}_r =
\begin{bmatrix}
\bm{0}    \\
2(\bm{I}_{nq}\otimes\bm{a}_r)\widetilde{\bm{D}}      
\end{bmatrix},
\]
where $\bm{I}_{k}$ denotes the $k\times k$ identity matrix.
The bias vector
$\bm{B}(\widehat{\bm{\theta}})$ is simply the set coefficients from
the ordinary weighted lest-squares regression of the $\widetilde{\bm{\xi}}$
on the columns of $\widetilde{\bm{F}}$, using weights in $\widetilde{\bm{H}}$.
The bias vector $\bm{B}(\widehat{\bm{\theta}})$
will be small when $\widetilde{\bm{\xi}}$ is orthogonal to the columns
of $\widetilde{\bm{H}}\widetilde{\bm{F}}$
and it can be large when $n$ is small.
Note that equation~(\ref{BIAS}) involves simple operations on matrices and vectors
and we can calculate the bias $\bm{B}(\widehat{\bm{\theta}})$ numerically via software
with numerical linear algebra facilities such as {\tt Ox} \citep{DcK2006} and {\tt R}
\citep{R2006} with minimal effort.

After some algebra, we have
\[
\bm{a}_r =  \bm{1}_n \otimes 
\begin{pmatrix}
\bm{F}_{\bm{\beta}_0}^{(r)}\\
\bm{0}
\end{pmatrix}, \quad
\bm{a}_s =  \bm{1}_n \otimes 
\begin{pmatrix}
\bm{F}_{\bm{\beta}_1}^{(s)}\bm{\mu}_{\bm{x}}\\
\bm{0}
\end{pmatrix},\quad
\bm{a}_t =  \bm{1}_n \otimes 
\begin{pmatrix}
\bm{\beta}_1\bm{F}_{\bm{\mu}_{\bm{x}}}^{(t)}\\
\bm{F}_{\bm{\mu}_{\bm{x}}}^{(t)}
\end{pmatrix}\quad{\rm and}\quad
\bm{a}_u = \bm{0},
\]
for $r = 1, \ldots, v$; $s = v+1, \ldots, v(m+1)$; $t = v(m+1)+1, \ldots, v(m+1) + m$;
and $u = (v+1)(m+1), \ldots, p$; where $p = v(m+1) + m + m(m+1)/2 + v(v+1)/2$.
(Here, $\bm{1}_{n}$ denotes an $n\times 1$ vector of ones.) Moreover,
\[
\bm{a}_{rs} = \bm{1}_n \otimes 
\begin{pmatrix}
\bm{F}_{\bm{\beta}_1}^{(s)}\bm{F}_{\bm{\mu}_{\bm{x}}}^{(r)}\\
\bm{0}
\end{pmatrix},
\]
for all $r$ and $s$,  
\[
\bm{C}_{s} = \bm{I}_n \otimes 
\begin{pmatrix}
\bm{F}_{\bm{\beta}_1}^{(s)}\bm{\Sigma}_{\bm{x}}\bm{\beta}_1^{\top} + \bm{\beta}_1\bm{\Sigma}_{\bm{x}}
\bm{F}_{\bm{\beta}_1}^{(s)\top}  & \bm{F}_{\bm{\beta}_1}^{(s)}\bm{\Sigma}_{\bm{x}}\\
\bm{F}_{\bm{\beta}_1}^{(s)}\bm{\Sigma}_{\bm{x}} & \bm{0}
\end{pmatrix}, \
\bm{C}_{t} = \bm{I}_n \otimes 
\begin{pmatrix}
\bm{\beta}_1\bm{F}_{\bm{\Sigma}_{\bm{x}}}^{(t)}\bm{\beta}_1^{\top}  &
\bm{\beta}_1\bm{F}_{\bm{\Sigma}_{\bm{x}}}^{(t)}\\
\bm{F}_{\bm{\Sigma}_{\bm{x}}}^{(t)}\bm{\beta}_1^{\top}  & \bm{0}
\end{pmatrix}
\]
and
\[
\bm{C}_{u} = \bm{I}_n \otimes 
\begin{pmatrix}
\bm{F}_{\bm{\Sigma}_{\bm{q}}}^{(u)} & \bm{0}\\
\bm{0}  & \bm{0}
\end{pmatrix},
\] 
for $s = v+1, \ldots, v(m+1)$; $t = v(m+1)+1, \ldots, v(m+1) + m$; and $u = (v+1)(m+1), \ldots, p$.
Additionally,
\[
\bm{C}_{rs} = \bm{I}_n \otimes 
\begin{pmatrix}
\bm{F}_{\bm{\beta}_1}^{(s)}\bm{\Sigma}_{\bm{x}}\bm{F}_{\bm{\beta}_1}^{(r)\top} +
\bm{F}_{\bm{\beta}_1}^{(r)}\bm{\Sigma}_{\bm{x}} \bm{F}_{\bm{\beta}_1}^{(s)\top}  & \bm{0}\\
\bm{0} & \bm{0}
\end{pmatrix}
\] 
and
\[
\bm{C}_{tu} = \bm{I}_n \otimes 
\begin{pmatrix}
\bm{F}_{\bm{\beta}_1}^{(u)}\bm{F}_{\bm{\Sigma}_{\bm{x}}}^{(t)}\bm{\beta}_1^{\top} +
\bm{\beta}_1\bm{F}_{\bm{\Sigma}_{\bm{x}}}^{(t)}\bm{F}_{\bm{\beta}_1}^{(s)\top} &
\bm{F}_{\bm{\beta}_1}^{(u)}\bm{F}_{\bm{\Sigma}_{\bm{x}}}^{(t)}\\
\bm{F}_{\bm{\Sigma}_{\bm{x}}}^{(t)}\bm{F}_{\bm{\beta}_1}^{(u)\top}  & \bm{0}
\end{pmatrix},
\]
for $r,s,u = v+1, \ldots, v(m+1)$; $t = v(m+1)+1, \ldots, v(m+1) + m$; and $\bm{C}_{rs} = \bm{0}$ otherwise.

Therefore, in the measurement error model defined by the equations (\ref{Mainmodel}) and
(\ref{model}), all quantities necessary to compute the $O(n^{-1})$
bias of $\widehat{\bm{\theta}}$ using expression (\ref{BIAS}) are given.
On the right-hand side of expression~(\ref{BIAS}),
consistent estimates of the parameter $\bm{\theta}$ can be inserted to define
the corrected MLE $\widetilde{\bm{\theta}} = \widehat{\bm{\theta}} - \widehat{\bm{B}}(\widehat{\bm{\theta}})$,
where $\widehat{\bm{B}}(\cdot)$ denotes the MLE of $\bm{B}(\cdot)$,
that is, the unknown parameters are replaced by their MLEs.
The bias-corrected estimate (BCE) $\widetilde{\bm{\theta}}$  is expected to have
better sampling properties than the uncorrected estimator, $\widehat{\bm{\theta}}$.
In fact, we present some simulations in Section~\ref{simulations} to show that
$\widetilde{\bm{\theta}}$ has smaller bias than its corresponding MLE,
thus suggesting that the bias corrections have the effect of shifting the modified estimates
toward to the true parameter values.

The BCEs can always be defined if the joint cumulants of the derivatives of the log-likelihood 
function and the MLEs exist.
Although, in some situations (for example, homoskedastic simple errors-in-variables 
model), the first moment of the MLEs is not defined, it is still 
possible to define such ``corrected'' estimators from $\bm{B}(\widehat{\bm{\theta}})$.
 In this case, the interpretation of $\bm{B}(\widehat{\bm{\theta}})$ may not be the second-order bias of
 $\widehat{\bm{\theta}}$, but it is still being an ``adjustement'' factor of the location of the MLEs.
Patriota and Lemonte (2009) present some simulation studies considering a simple linear errors-in-variables
model in which is showed that the BCEs have better performance than the MLEs for finite sample sizes.
In general, it is very hard to verify if the MLEs of the parameters of the model considered in this paper
have defined expectations, but the simulation studies presented in Section~\ref{simulations}
indicate a better performance of the corrected 
estimators than the uncorrected ones and, therefore, we advise to use the corrected estimators.

\section{Biases of the MLEs $\widehat{\bm{\mu}}_{i}$ and $\widehat{\bm{\Sigma}}_{i}$}\label{mu-sigma-vies}

In this section, we give matrix formulae for the $O(n^{-1})$ biases of the MLEs of the
$i$th mean $\bm{\mu}_{i}=\bm{\mu}_{i}(\bm{\theta})$
and $i$th variance-covariance vector $\bm{\Sigma}_{i}^{*} = \mbox{vech}(\bm{\Sigma}_{i}(\bm{\theta}))$.
Let $q_{1} = v+m$ and $q_{2} = q_{1}(q_{1} + 1)/2$. Additionally, let
$\bm{A} = [\bm{A}_1, \ldots, \bm{A}_n]^{\top}$
be a $np \times p$ matrix, where $\bm{A}_i$ is a $p \times p$ matrix, then we define 
$\mbox{tr}^{*}(\bm{A}) = [\mbox{tr}(\bm{A}_1), \ldots,\mbox{tr}(\bm{A}_n)]^{\top}$.

From a Taylor series expansion of $\widehat{\bm{\mu}}_{i} = \bm{\mu}_{i}(\widehat{\bm{\theta}})$,
we obtain up to an error of order $O(n^{-2})$: 
\[
\bm{B}(\widehat{\bm{\mu}}_{i}) = \bm{L}_{i}\bm{B}(\widehat{\bm{\theta}})+\frac{1}{2}\mbox{tr}^{*}[\bm{M}_i
\textrm{Cov}(\widehat{\bm{\theta}})],
\]
where $\bm{L}_{i}$ is a $q_{1}\times p$ matrix of first partial derivatives $\partial\bm{\mu}_{i}/
\partial\theta_{r}$ (for $r =1,2,\ldots,p$),
$\bm{M}_i = [\bm{M}_{i1},\ldots,\bm{M}_{iq_{1}}]^{\top}$ is a
$q_{1}p\times p$ matrix of second partial derivatives, where $\bm{M}_{il}$
is a $p\times p$ matrix with elements $\partial^{2}\mu_{il}/\partial\theta_{r}\partial\theta_{s}$
(for $r,s=1,\ldots,p$ and $l = 1,2,\ldots,q_{1}$),
$\textrm{Cov}(\widehat{\bm{\theta}})=\bm{K}_{\bm{\theta}}^{-1}$ 
is the asymptotic covariance matrix of $\widehat{\bm{\theta}}$ and the vector 
$\bm{B}(\widehat{\bm{\theta}})$ was defined before. 
All quantities in the above equation should be evaluated at $\widehat{\bm{\theta}}$.
The asymptotic variance of $\widehat{\bm{\mu}}_{i}$ can also be expressed explicitly 
in terms of the covariance of $\widehat{\bm{\theta}}$ by
\[
\textrm{Var}(\widehat{\bm{\mu}}_{i})=\bm{L}_{i}\textrm{Cov}(\widehat{\bm{\theta}})\bm{L}_{i}^{\top}.
\]

The second-order bias of $\widehat{\bm{\Sigma}}_{i}^{*}$ is obtained by expanding
$\widehat{\bm{\Sigma}}_{i}^{*} = \bm{\Sigma}_{i}^{*}(\widehat{\bm{\theta}})$ in Taylor
series. Then, the $O(n^{-1})$ bias of $\widehat{\bm{\Sigma}}_{i}^{*}$ is written as:
\[
\bm{B}(\widehat{\bm{\Sigma}}_{i}^{*}) = \bm{L}_{i}^{*}\bm{B}(\widehat{\bm{\theta}})+
\frac{1}{2}\mbox{tr}^{*}[\bm{M}_i^{*}\textrm{Cov}(\widehat{\bm{\theta}})],
\]
where $\bm{L}_{i}^{*}$ is a $q_{2}\times p$ matrix of first partial derivatives $\partial\bm{\Sigma}_{i}^{*}/
\partial\theta_{r}$ (for $r =1,2,\ldots,p$),
$\bm{M}_i^{*} = [\bm{M}_{i1}^{*},\ldots,\bm{M}_{iq_{2}}^{*}]^{\top}$ is a
$q_{2}p\times p$ matrix of second partial derivatives, where $\bm{M}_{il}^{*}$
is a $p\times p$ matrix with elements $\partial^{2}\Sigma_{il}^{*}/\partial\theta_{r}\partial\theta_{s}$
(for $r,s=1,\ldots,p$ and $l = 1,2,\ldots,q_{2}$).

Therefore, we are now able to define the following second-order bias-corrected estimators for
$\widehat{\bm{\mu}}_{i}$ and $\widehat{\bm{\Sigma}}_{i}^{*}$:
\[
\widetilde{\bm{\mu}}_{i} = \widehat{\bm{\mu}}_{i} - \widehat{\bm{B}}(\widehat{\bm{\mu}}_{i})
\quad{\rm and}\quad
\widetilde{\bm{\Sigma}}_{i}^{*} = \widehat{\bm{\Sigma}}_{i}^{*} - \widehat{\bm{B}}(\widehat{\bm{\Sigma}}_{i}^{*}).
\]

It is clear that the $O(n^{-1})$ bias of any other function of $\bm{\theta}$,
say $\bm{\Psi}(\bm{\theta})$ ($h\times 1$), can be obtained easily by Taylor series expansion: 
\[
\bm{B}(\widehat{\bm{\Psi}}) = \bm{\nabla}_{\bm{\Psi}}^{(1)}\bm{B}(\widehat{\bm{\theta}})+
\frac{1}{2}\mbox{tr}^*[\bm{\nabla}_{\bm{\Psi}}^{(2)}\textrm{Cov}(\widehat{\bm{\theta}})],
\]
where $\bm{\nabla}_{\bm{\Psi}}^{(1)}$ is a $h\times p$ matrix of first partial derivatives $\partial\bm{\Psi}/
\partial\theta_{r}$ (for $r =1,2,\ldots,p$) and
$\bm{\nabla}_{\bm{\Psi}}^{(2)} = [\bm{\nabla}_{\bm{\Psi}1}^{(2)},\ldots,\bm{\nabla}_{\bm{\Psi}h}^{(2)}]^{\top}$ is a
$hp\times p$ matrix of second partial derivatives, where $\bm{\nabla}_{\bm{\Psi}l}^{(2)}$
is a $p\times p$ matrix with elements $\partial^{2}\Psi_{l}/\partial\theta_{r}\partial\theta_{s}$
(for $r,s=1,\ldots,p$ and $l = 1,2,\ldots,h$).

\section{Numerical results}\label{simulations}

We shall use Monte Carlo simulation to evaluate the finite sample performance of the MLEs
attained using the iterative formula~(\ref{Fisher-Scoring}) and of their corresponding
bias-corrected versions for a heteroskedastic errors-in-variables model presented in~(\ref{model})
with $m=v=1$. The sample sizes considered
were $n = 40, 60, 100$ and 200, the number of Monte Carlo replications was 10,000.
All simulations were performed using the {\tt R} programming language \citep{R2006}.

We consider the simple errors-in-variables model
\begin{equation*}\label{model-simple}
Y_i = y_i + \eta_{y_{i}}\quad{\rm and}\quad X_i = x_i + \eta_{x_i},
\end{equation*}
with $y_i | x_i\stackrel{ind}{\sim}\mathcal{N}(\beta_0 + \beta_1 x_i,\sigma^2)$.
This model was studied by \cite{Kulathinal2002}. The errors $\eta_{y_i}$ and $\eta_{x_i}$
are independent of the unobservable covariate $x_i$ and are distributed as
\[
\begin{pmatrix}
 \eta_{y_i}\\
 \eta_{x_i}\\
\end{pmatrix}
\stackrel{ind}{\sim}\mathcal{N}_2
\begin{bmatrix}
\begin{pmatrix}
 0\\
 0\\
\end{pmatrix},
\begin{pmatrix}
 \tau_{y_i} & 0        \\
 0         & \tau_{x_i}\\
\end{pmatrix}\end{bmatrix},
\]
where the variances $\tau_{y_i}$ and $\tau_{x_i}$ are known for all $i=1,\ldots,n$. 
Supposing in addition that $x_i\stackrel{iid}{\sim}
\mathcal{N}(\mu_x,\sigma^2_x)$, we have that the joint distribution of
the observed variables can be expressed as
\begin{equation*}\label{model_Dist}
\begin{pmatrix}
Y_i\\
X_i\end{pmatrix}
\stackrel{ind}{\sim}\mathcal{N}_2
\begin{bmatrix}
\begin{pmatrix}
\beta_0 + \beta_1 \mu_x\\
\mu_x
\end{pmatrix},
\begin{pmatrix}
\beta_1^2\sigma_x^2 + \tau_{y_i} + \sigma^2& \beta_1 \sigma_x^2\\
\beta_1  \sigma_x^2 &\sigma_x^2 + \tau_{x_i}
\end{pmatrix}
\end{bmatrix}.
\end{equation*}
Define $\bm{\theta} = (\beta_0, \beta_1, \mu_x, \sigma_x^2, \sigma^2)^{\top}$, 
\[
\bm{\mu}_i (\bm{\theta})= {\beta_0 + \beta_1\mu_x \choose \mu_x}
\quad \mbox{and} \quad
\bm{\Sigma}_i(\bm{\theta}) =
\begin{pmatrix}
\beta_1^2\sigma_x^2 + \sigma^2 + \tau_{y_i} & \beta_1 \sigma_x^2 \\
\beta_1\sigma_x^2 & \sigma_x^2+ \tau_{x_i}
\end{pmatrix}.
\]
From the previous expressions, we have immediately that
\[
\bm{a}_1 =  \bm{1}_n \otimes 
\begin{pmatrix}
1\\
0
\end{pmatrix}, \quad
\bm{a}_2 =  \bm{1}_n \otimes 
\begin{pmatrix}
\mu_x\\
0
\end{pmatrix}, \quad
\bm{a}_3 =  \bm{1}_n \otimes 
\begin{pmatrix}
\beta_{1}\\
1
\end{pmatrix}, \quad
\bm{a}_4 = \bm{a}_5 = \bm{0}
\]
and $\bm{a}_{rs} = \bm{0}$ for all $r, s$ except for
\[
\bm{a}_{23} = \bm{a}_{32} =  \bm{1}_n \otimes 
\begin{pmatrix}
1\\
0
\end{pmatrix}.
\]
Also, $\bm{C}_{1} = \bm{C}_{3} = \bm{0}$ and
\[
\bm{C}_{2} = \bm{I}_n \otimes 
\begin{pmatrix}
2\beta_1\sigma_x^2 & \sigma_x^2\\
\sigma_x^2  &0 
\end{pmatrix}, \
\bm{C}_{4} = \bm{I}_n \otimes 
\begin{pmatrix}
\beta_1^2 & \beta_1\\
\beta_1  & 1
\end{pmatrix}\quad\mbox{and}\quad
\bm{C}_{5} = \bm{I}_n \otimes 
\begin{pmatrix}
1 & 0\\
0  & 0
\end{pmatrix}.
\]
Additionally, $\bm{C}_{rs} = \bm{0}$ for all $r, s$  except for
\[
\bm{C}_{22} = \bm{I}_n \otimes 
\begin{pmatrix}
2\sigma_x^2 & 0\\
0  & 0
\end{pmatrix} \ \mbox{and} \ 
\bm{C}_{24} = \bm{C}_{42} = \bm{I}_n \otimes 
\begin{pmatrix}
2\beta_1 & 1\\
1  & 0
\end{pmatrix}.
\]
Thus, $\widetilde{\bm{D}} = (\bm{a}_1, \bm{a}_2, \bm{a}_3,\bm{0},\bm{0})$ and
$\widetilde{\bm{V}} = (\bm{0}, \vvec(\bm{C}_2), \bm{0},\vvec(\bm{C}_4),\vvec(\bm{C}_5))$. 
Therefore, all the quantities necessary to calculate $\bm{B}(\widehat{\bm{\theta}})$ using
expression~(\ref{BIAS}) are given.

In order to analyze the point estimation results, we computed, for each sample size
and for each estimator: relative bias (the relative bias of an estimator $\widehat{\theta}$
is defined as $\{\Es(\widehat{\theta}) - \theta\}/\theta$, its estimate being obtained
by estimating $\Es(\widehat{\theta})$ by Monte Carlo) and root mean square error, i.e.,
$\sqrt{{\rm MSE}}$, where MSE is the mean squared error estimated
from the 10,000 Monte Carlo replications.
For practical reasons and without loss of generality, we adopt the same setting of
parameters chosen by \cite{deCastro2008}. (The parameters are the MLEs for the model
parameters using a real data set presented in the next section.) We take $\beta_0=-2$,
$\beta_1=0.5$, $\mu_x=-2$, $\sigma_x^2=4$ and $\sigma^2 = 10$. We also consider
two types of heteroskedasticity as studied by \cite{Patriota},
namely: (a) $\sqrt{\tau_{x_i}} \sim U(0.5,1.5)$ and
$\sqrt{\tau_{y_i}} \sim U(0.5,4)$,  where $U(a,b)$ means uniform distribution on $[a,b]$;
(b) $\sqrt{\tau_{x_i}} = 0.1|x_i|$ and  $\sqrt{\tau_{y_i}} = 0.1|-2 + 0.51x_i|$, i.e.,
the variances depend on the unknown covariate. We remark that the variances are
considered to be known and kept fixed in all Monte Carlo simulations.

Table~\ref{tab:1} shows simulation results for an errors-in-variables model with
a uniform heteroskedasticity. The figures in this table reveal that the maximum-likelihood 
estimators of the parameters can be substantially biased when the
sample size is small, and that the bias correction we derived in the previous section
is very effective. For instance, when $n = 40$ the biases of the estimators of $\beta_0$,
$\beta_1$, $\mu_{x}$, $\sigma_{x}^2$ and $\sigma^2$ average $-0.02244$ whereas
the biases of the corresponding bias-adjusted estimators average $-0.00276$;
that is, the average bias (in value absolute) of the MLEs is almost ten times greater
than that of the corrected estimators. In particular, the maximum-likelihood estimators
of $\sigma_{x}^2$ and $\sigma^2$ display substantial bias, and the bias correction
proves to be quite effective when applied to these estimators.

\begin{table}[htp]\renewcommand{\arraystretch}{1.1}
\caption{Relative bias and $\sqrt{{\rm MSE}}$ of uncorrected and corrected 
estimates with a uniform heteroskedasticity: 
$\sqrt{\tau_{x_i}} \sim U(0.5,1.5)$ and  $\sqrt{\tau_{y_i}} \sim U(0.5,4)$.}\label{tab:1}
\begin{tabular}{ccrrrrr}\hline
 &       & \multicolumn{2}{c}{MLE} & &\multicolumn{2}{c}{BCE}\\\cline{3-4}\cline{6-7}                                                         
$n$ &  $\bm{\theta}$ &Rel.~bias & $\sqrt{{\rm MSE}}$ && Rel.~bias & $\sqrt{{\rm MSE}}$  \\\hline  
40& $\beta_0$        &  $-0.0173$  & 0.99      &&   $-0.0043$       &  0.97    \\               
 &  $\beta_1$        &  $ 0.0315$  & 0.38      &&   $ 0.0054$       &  0.37    \\            
 &  $\mu_{x}$        &  $-0.0018$  & 0.35      &&   $-0.0018$       &  0.35    \\       
 &  $\sigma_{x}^2$   &  $-0.0351$  & 1.11      &&   $-0.0045$       &  1.13    \\        
 &  $\sigma^2$       &  $-0.0895$  & 3.31      &&   $-0.0086$       &  3.38    \\\hline 
60 &  $\beta_0$      &  $-0.0139$  & 0.77      &&   $-0.0061$       &  0.76  \\      
 &  $\beta_1$        &  $ 0.0213$  & 0.29      &&   $ 0.0058$       &  0.29  \\   
 &  $\mu_{x}$        &  $ 0.0009$  & 0.28      &&   $ 0.0009$       &  0.28  \\   
 &  $\sigma_{x}^2$   &  $-0.0239$  & 0.89      &&   $-0.0036$       &  0.90  \\   
 &  $\sigma^2$       &  $-0.0548$  & 2.60      &&   $-0.0018$       &  2.64  \\\hline  
100 &$\beta_0$       &  $-0.0100$  & 0.68      &&   $-0.0037$       &  0.67  \\   
 &  $\beta_1$        &  $ 0.0168$  & 0.26      &&   $ 0.0042$       &  0.25  \\ 
 &  $\mu_{x}$        &  $ 0.0001$  & 0.25      &&   $ 0.0001$       &  0.25  \\
 &  $\sigma_{x}^2$   &  $-0.0135$  & 0.80      &&   $ 0.0022$       &  0.81  \\
 &  $\sigma^2$       &  $-0.0424$  & 2.40      &&   $ 0.0003$       &  2.43  \\\hline 
200 &  $\beta_0$     &  $-0.0049$  & 0.59      &&   $-0.0006$       &  0.59  \\      
 &  $\beta_1$        &  $ 0.0127$  & 0.22      &&   $ 0.0041$       &  0.22  \\     
 &  $\mu_{x}$        &  $ 0.0013$  & 0.23      &&   $ 0.0013$       &  0.23  \\    
 &  $\sigma_{x}^2$   &  $-0.0116$  & 0.70      &&   $ 0.0008$       &  0.70  \\   
 &  $\sigma^2$       &  $-0.0350$  & 2.09      &&   $-0.0014$       &  2.11  \\\hline 
\multicolumn{6}{l}{BCE: bias-corrected estimator.}                                  
\end{tabular}                                                                                                            \end{table}                                                                                                              

Table~\ref{tab:2} displays simulation results for an errors-in-variables model with
a nonuniform heteroskedasticity. We note that the bias-adjusted estimator again
displays smaller bias than the standard maximum-likelihood estimator. This suggests
that the second-order bias of MLEs should not be ignored in samples of small to moderate
sizes since they can be nonnegligible. Note also that root mean square error decrease
with $n$, as expected. Additionally, we note that all estimators have similar root mean squared errors. 

It is interesting to note that the finite-sample performance of the estimator
of $\sigma_{x}^{2}$ deteriorate when we pass from the model with a uniform
heteroskedasticity to the model with a nonuniform heteroskedasticity
(see Tables 1 and 2). For instance, when $n=100$, the relative biases
of $\widehat{\sigma}_{x}^{2}$ (MLE) were $-0.0135$ (uniform heteroskedasticity)
and $-0.0484$ (nonuniform heteroskedasticity), which amounts to an increase
in relative biases of nearly $3.5$ times.

\begin{table}[htp]\renewcommand{\arraystretch}{1.1}
\caption{Relative bias and $\sqrt{{\rm MSE}}$ of uncorrected and corrected    
estimates with a nonuniform heteroskedasticity:                                                        
$\sqrt{\tau_{x_i}} = 0.1|x_i|$ and $\sqrt{\tau_{y_i}} = 0.1|\beta_0 + \beta_1x_i|$. }\label{tab:2}                                       
\begin{tabular}{ccrrrrr}\hline                                                                                                     
 &       & \multicolumn{2}{c}{MLE} & &\multicolumn{2}{c}{BCE}\\\cline{3-4}\cline{6-7} 
$n$ &  $\bm{\theta}$ &Rel.~bias & $\sqrt{{\rm MSE}}$ && Rel.~bias & $\sqrt{{\rm MSE}}$  \\\hline 
40& $\beta_0$        &  $-0.0026$  &0.73          &&   $-0.0018$       &  0.73    \\  
 &  $\beta_1$        &  $ 0.0292$  &0.27          &&   $ 0.0276$       &  0.27    \\ 
 &  $\mu_{x}$        &  $-0.0228$  &0.32          &&   $-0.0228$       &  0.32    \\ 
 &  $\sigma_{x}^2$   &  $-0.0594$  &0.91          &&   $-0.0354$       &  0.92    \\ 
 &  $\sigma^2$       &  $-0.0540$  &2.26          &&   $-0.0056$       &  2.30    \\\hline
60 & $\beta_0$       &  $ 0.0008$  &0.59          &&   $ 0.0013$       &  0.59  \\ 
 &  $\beta_1$        &  $ 0.0203$  &0.22          &&   $ 0.0192$       &  0.22  \\ 
 &  $\mu_{x}$        &  $-0.0208$  &0.26          &&   $-0.0208$       &  0.26  \\ 
 &  $\sigma_{x}^2$   &  $-0.0502$  &0.76          &&   $-0.0340$       &  0.75  \\ 
 &  $\sigma^2$       &  $-0.0332$  &1.85          &&   $-0.0002$       &  1.88  \\\hline
100 &  $\beta_0$     &  $ 0.0013$  &0.51          &&   $ 0.0016$       &  0.51  \\      
 &  $\beta_1$        &  $ 0.0184$  &0.19          &&   $ 0.0176$       &  0.19  \\      
 &  $\mu_{x}$        &  $-0.0198$  &0.23          &&   $-0.0198$       &  0.23  \\      
 &  $\sigma_{x}^2$   &  $-0.0484$  &0.65          &&   $ -0.0363$      &  0.65  \\      
 &  $\sigma^2$       &  $-0.0223$  &1.61          &&   $ 0.0027 $      &  1.64  \\\hline
200 &  $\beta_0$     &  $ 0.0036$  &0.45          &&   $ 0.0039 $      &  0.45  \\     
 &  $\beta_1$        &  $ 0.0165$  &0.17          &&   $ 0.0159 $      &  0.17  \\     
 &  $\mu_{x}$        &  $-0.0186$  &0.20          &&   $-0.0186 $      &  0.20  \\     
 &  $\sigma_{x}^2$   &  $-0.0474$  &0.59          &&   $-0.0377 $      &  0.58  \\     
 &  $\sigma^2$       &  $-0.0204$  &1.41          &&   $-0.0004 $      &  1.43  \\\hline 
\multicolumn{6}{l}{BCE: bias-corrected estimator.}                                      
\end{tabular}                                                                          
\end{table}                                                                          

\section{Application}\label{application}                                                            

We shall now present an application of the model described in Section~\ref{modelo-heteros} where $v=m=1$.
We analyze a epidemiological data set from
the WHO MONICA (World Health Organization Multinational MONitoring of trends and
determinants in CArdiovascular disease) Project. This data set was previously studied by 
\cite{Kulathinal2002} and \cite{deCastro2008} where the ML approach was adopted to estimate 
the model parameters.

The main goal of this project is
to monitor trends in cardiovascular diseases and relate it with known risk
factors. Here, $y$ is the trends in cardiovascular mortality 
and coronary heart disease and $x$ is the changes in known risk factors. 
The risk score was defined as a linear combination of smoking status, systolic 
blood pressure, body mass index and total cholesterol level. Note that, these
variables are non-observable indexes therefore they need to be estimated in some 
way. Follow up studies where conducted using proportional hazards models which can 
provide the observed ($Y$ and $X$) indexes and the measurement error variances. 

The latent variables $y$ and $x$ are linearly related as 
\[
y_i = \beta_0 + \beta_1 x_i + q_i,\quad i = 1, \ldots, n.
\]
As the variables $y_i$ and $x_i$ are not directly observable,
surrogate variables $Y_i$ and $X_i$ are observed in their place, respectively. Such surrogate
variables are attained from an analytical treatment of the data collection process.
The data set are divided into two groups, namely: men ($n=38$) and women ($n=36$).

In what follows, we compare the MLEs with the bias-corrected estimators. 
Table \ref{applic} presents the MLEs, its standard deviation, its second-order biases
and the corrected estimates. It can be seen that, the greater is the
standard deviation of the MLE, the more distant from zero is its respectively second-order
bias. As concluded in the simulation studies, the biases of the variances estimates
are larger than of those produced by the line estimators. The second-order biases of
the MLEs can be expressed as a percentage of the MLEs. That is, for the men data set,
the second-order biases are $-0.21\%$, 0.85\%, 0.00\%, $-2.92\%$ and $-9.21\%$ of the total
amount of the MLEs of $\beta_0$, $\beta_1$, $\mu_x$, $\sigma_x^2$ and $\sigma^2$,
respectively. For the women data set, the second-order biases are 52.96\%,
1.21\%, 0.00\%, $-3.16\%$ and $-10.19\%$ of the MLEs of $\beta_0$, $\beta_1$, $\mu_x$,
$\sigma_x^2$ and $\sigma^2$, respectively. It shows that the second-order biases
of the MLEs are more pronounced in the women data set, mainly for the intercept estimator.
\begin{table}[htp]\renewcommand{\arraystretch}{1.1}
\caption{MLEs and bias-corrected estimates.}\label{applic}
\begin{tabular}{ccrrrr}\hline
&Parameter       &  MLEs     &  S.E.   &   Bias    &   BCEs        \\\hline
&$\beta_0$       & $-2.0799$ & 0.5285  & $ 0.0044$&    $-2.0843$ \\
&$\beta_1$       & $0.4690$  & 0.2339  & $ 0.0040$&    $0.4650$ \\
Men&$\mu_{x}$    & $-1.0924$ & 0.3550  & $ 0.0000$&    $-1.0924$ \\
&$\sigma_{x}^2$  & $4.3163$  & 1.0969  & $ -0.1261$&   $4.4423$ \\
&$\sigma^2$      & $4.8883$  & 1.7790  & $ -0.4501$&   $5.3384$ \\\hline
&Parameter       &  MLEs     &  S.E.   &   Bias    &   BCEs        \\\hline
&$\beta_0$       & $0.0321$  & 1.1121  & $ 0.0170$&    $0.0151$ \\
&$\beta_1$       & $0.6790$  & 0.4072  & $ 0.0082$&    $0.6708$ \\
Women&$\mu_{x}$  & $-2.0677$ & 0.3386  & $ 0.0000$&    $-2.0677$ \\
&$\sigma_{x}^2$  & $3.6243$  & 0.9695  & $ -0.1146$&   $3.7389$ \\
&$\sigma^2$      & $11.0809$ & 4.2425  & $ -1.1289$&   $12.2098$ \\\hline
\multicolumn{5}{l}{BCE: bias-corrected estimates.}                                              
\end{tabular}                                                                 
\end{table}
      
\section{Conclusions}\label{conclusion}

We derive a bias-adjustment scheme to eliminate the second-order
biases of the max\-imum-likelihood estimates in a heteroskedastic multivariate
errors-in-variables regression model using the general matrix formulae
for the second-order bias derived by \cite{PatriotaLemonte2009}. 
The simulation results presented show that the MLEs can be considerably biased. The
bias correction derived in this paper is very effective, even when the
sample size is large. Indeed, the bias correction mechanism adopted yields
modified maximum-likelihood estimates which are nearly unbiased. 
Additionally, many errors-in-variables models are special cases of the
proposed model and the results obtained here can be easily particularized to these submodels.
We also present an application to a real data set.

\section*{Acknowledgments}

This work was partially supported by FAPESP (Funda\c c\~ao de Amparo \`a Pesquisa do Estado de S\~ao Paulo, 
Brazil) and CNPq (Conselho Nacional de Desenvolvimento Cient\'ifico e Tecnol\'ogico, Brazil). The
authors thank Dr. Kari Kuulasmaa (National Public Health Institute, Finland) for kindly supplying the 
data of our application. The authors are also grateful to a referee for helpful comments and
suggestions.

\end{document}